\documentclass[twocolumn,showpacs,preprintnumbers,amsmath,amssymb]{revtex4}
\usepackage{amssymb}
\usepackage[dvips]{graphicx}
\begin{document}

\title{How normal is the "normal" state of superconducting cuprates?}
\author{V. N. Zavaritsky and A. S. Alexandrov}

\affiliation{Department of Physics, Loughborough University,
Loughborough LE11 3TU, United Kingdom}

\begin{abstract}
High magnetic field studies of the cuprate superconductors revealed    a non-BCS  temperature
 dependence of the upper critical field $H_{c2}(T)$ determined resistively by several groups. These determinations caused some doubts on the grounds of the contrasting effect of the magnetic field on the in-plane, $\rho_{ab}$, and out-of-plane, $\rho_{c}$ resistances reported for large sample of Bi2212.   Here we present careful  measurements of both $\rho_{ab}(B)$ and $\rho_{c}(B)$ of  tiny Bi2212 crystals in  magnetic fields up to 50 Tesla. None of our measurements revealed a situation when on field increase $\rho_c$ reaches its maximum while $\rho_{ab}$ remains very small if not zero. The resistive $H_{c2}(T)$ estimated from $\rho_{ab}(B)$ and $\rho_{c}(B)$ are
approximately the same. We also present a simple explanation of
the unusual Nernst signal in superconducting cuprates as a normal
state phenomenon. Our results support any theory of cuprates,
which describes the state above the resistive phase transition as
perfectly 'normal' with a zero off-diagonal order parameter.
\end{abstract}


\maketitle

A  pseudogap is believed to be responsible for the non Fermi-liquid normal state of cuprate superconductors. Various
microscopic models of the pseudogap proposed are mostly based on
the strong electron correlations\cite{tim}, and/or  on the strong
electron-phonon interaction\cite{alebook}. There is also a
phenomenological scenario\cite{Kiv}, where the superconducting
order parameter (the Bogoliubov-Gor'kov anomalous average
$F(\mathbf{r,r^{\prime}})=\langle \psi _{\downarrow
}(\mathbf{{r})\psi _{\uparrow}({r^{\prime }}\rangle }$) does not
disappear at the resistive $T_{c}$ but at much higher (pseudogap)
temperature $T^{\ast }$. While the scenario \cite{Kiv} was found to
be inconsistent with the  `intrinsic tunnelling' I-V characteristics, the discovery of the Joule heating origin
 of the gap-like I-V nonlinearities made the
objection irrelevant \cite{ije-phys}.

In line with this scenario several authors\cite{mor,xu} suggested a radical revision of the magnetic phase diagram
of the cuprates with an upper critical field much higher than the resistive $H_{c2}(T)$-line.
In particular,  Ref.\cite{mor} questioned the resistive determination of $H_{c2}(T)$
 \cite{alezav,zav} claiming that while $\rho_c$ is a measure of the inter-plane tunnelling,
 only the in-plane data may represent a true normal state. The main argument
 in favour of this conclusion came from the radically different field dependencies of $\rho_c$ and $\rho_{ab}$ in Fig.2 of Ref.\cite{mor}, also shown in  our Fig.1 (inset B). According to these findings, magnetic field sufficient to recover normal state $\rho_c$, leaves in-plane superconductivity virtually unaffected. The difference suggests that Bi2212 crystals do not loose their  off-diagonal order in the Cu0$_2$ planes even well above $H_{c2}(T)$ determined from the c-axis data. This conclusion is based on one measurement so that it certainly deserves experimental verification, which was not possible until recently because of the lack of reliable $\rho_{ab}(B,T)$ for Bi2212. 

Quite similar conclusion followed from the thermomagnetic studies
of superconducting cuprates. A large Nernst signal  \emph{well
above}  $T_{c}$  has been attributed to a \emph{vortex} motion in
a number of cuprates \cite{xu,wang1}. As a result the magnetic
phase diagram of the cuprates has been revised with the upper
critical field $H_{c2}(T)$ curve not ending at $T_{c0}$ but at
much higher temperatures  \cite{wang1}. Most surprisingly,
Ref.\cite{wang1} estimated $H_{c2}$ \emph {at the zero-field
 transition temperature} of Bi2212, $T_{c0}$, as high as
50-150\,Tesla.

 On the other hand, any phase fluctuation  scenario such as of Ref. \cite{Kiv} is difficult to reconcile
 with the extremely sharp resistive and magnetic transitions at $T_{c}$ in single crystals of cuprates.
 Above $T_c$ the uniform magnetic susceptibility is paramagnetic and the resistivity is perfectly 'normal',
  showing only a few percent positive or negative magnetoresistance (MR). Both in-plane \cite{mac,boz,fra}
  and out-of-plane \cite {alezav} resistive transitions remain sharp in the magnetic field in
  high quality samples providing a reliable determination of a genuine $H_{c2}(T)$.
   These and some other observations \cite{lor}
  do not support any superconducting order parameter above $T_{c}$.

Resolution of these issues, which affect fundamental conclusions
about the nature of superconductivity in highly anisotropic
layered cuprates, requires further careful experiments and
transparent interpretations. Here we present systematic
measurements of both in-plane and out-of-plane MRs of small Bi2212
single crystals subjected to magnetic fields, $B\leq50$ Tesla,
$B\perp(ab)$. Our measurements reproduced neither the unusual
field dependence of $\rho_{ab}$  nor the contrasting effect of the
field as in Ref. \cite{mor}, which are most probably an
experimental artefact. On the contrary, they show that the
resistive upper critical fields estimated from the in-plane and
out-of-plane data are  nearly identical. We also present a simple
explanation of the unusual Nernst signal in cuprates as a normal
state phenomenon, thus supporting any microscopic theory of
cuprates with a zero off-diagonal order parameter above resistive
$T_c$.

Reliable measurements of the resistivity tensor require
defect-free samples. This is of prime importance for the in-plane
MR because even unit-cell scale defects will result in a
significant out-of-plane contribution owing to the extreme
anisotropy of Bi2212. Because of this reason much attention has
been paid to the sample preparation \cite{mos}. We studied
$\rho_c$ and $\rho_{ab}$ of the same high quality, optimally and
slightly underdoped Bi2212 crystals, T$_{c0}\approx$87-92$K$.
Differently from Ref. \cite{mor} small crystals were prepared in
order to reduce eddy currents and the forces acting on the sample
during the pulse. We measured $\rho_{c}$ on samples with in-plane
dimensions from $\simeq 30 \times 30 \mu m^2$ to $\simeq 80 \times
80 \mu m^2$ while $\rho_{ab}$ was studied on a longer crystals,
from $\simeq300\times11\mu m^2$ to $\simeq780\times22\mu m^2$.
Metallic type of zero-field $\rho_{ab}(T)$ and the sign of its
normal state MR \cite{mos} indicate vanishing out-of-plane
contribution. All samples selected for  $\rho_c$ and $\rho_{ab}$
measurements were cut from the same parent crystals of $1-3\mu m$
thickness. The absence of hysteresis in the $\rho(B)$ data
obtained on the rising and falling sides of the pulse and the
consistency of $\rho(B)$ taken at the same temperature in pulses
of different $B_{max}$ exclude any measurable heating effects.
Ohmic response is confirmed by a consistency of  the dc $\rho(B)$
measured at identical conditions with different currents,
$10$-$1000A/cm^2$ for $\rho_{ab}$ and $0.1$-$20A/cm^2$ for
$\rho_c$.

\begin{figure}
\begin{center}
\includegraphics[angle=-0,width=0.47\textwidth]{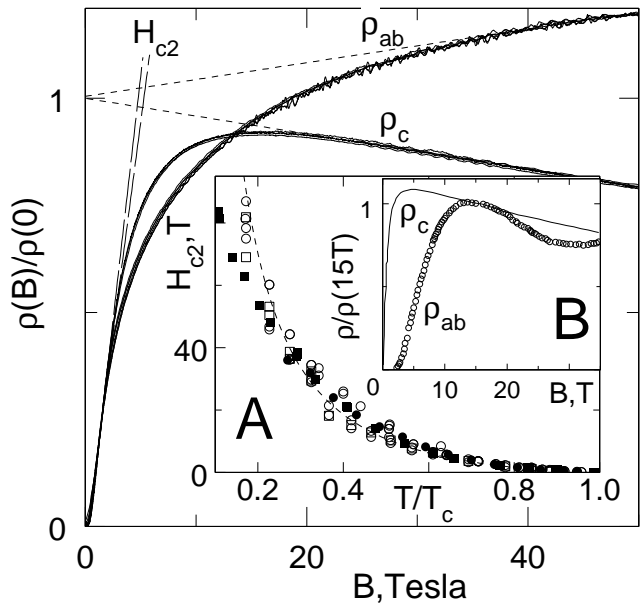}
\vskip -0.5mm \caption{ $\rho_c(B)$ and $\rho_{ab}(B)$ of Bi2212
normalised by corresponding $\rho_N(0,T)$ obtained with the linear
extrapolation from the normal state region (short dashes). The
linear fits, shown by long dashed lines, refer to the flux-flow
region. Inset A:   $H_{c2}$ estimated  from $\rho_{ab}(B)$ and
$\rho_{c}(B)$
 is shown by the open and  solid symbols respectively together with the fit,
 $H_{c2}(T) \sim (t^{-1}-t^{1/2})^{3/2}$, with $t=T/T_{c}$ \cite{ale} (dashed line).
Inset B shows $\rho_c$ and $\rho_{ab}$ from the inset to Fig.2 in Ref.\cite{mor}.
}
\end{center}
\end{figure}
Fig.1 shows the typical $\rho_c(B)$ and $\rho_{ab}(B)$  taken
below $T_{c0}$ of a Bi2212 single crystal. The low-field portions
of the curves correspond to the resistance driven by vortex
dynamics. Here a non-linear $\rho(B)$ dependence is followed by a
regime, where a linear dependence fits the experimental
observations rather well, Fig.1.  It is natural to attribute the
high field portions of the curves in Fig.1 (assumed to be above
H$_{c2}$) to a normal state. Then,  the c-axis high-field MR
appears to be negative and quasi-linear in B in a wide temperature
range both above and below $T_{c0}$. Contrary to  $\rho_{c}(B)$,
the normal state in-plane MR is {\it positive} (see \cite{mos} and
references therein for   an explanation). The reasonable
concordance of $H_{c2}(T)$ estimates from $\rho_{c}(B)$ and $\rho_{ab}(B)$ (Inset A to Fig.1) favours our association of the resistive $H_{c2}$ with the upper critical field especially given the apparently different mechanisms responsible for $\rho_{ab}$ and $\rho_{c}$ \cite{mos}.

Our conclusion is based on the results obtained during few hundred
measurements performed on three pairs of crystals. None of those
revealed a situation when on field increase $\rho_c$ reaches its
maximum while $\rho_{ab}$ remains very small if not zero as
reported in Ref. \cite{mor} (see  inset B in Fig.1). Since
 the authors of Ref.\cite{mor}  measured $'\rho_{ab} (B)'$ by
means of contacts situated on the same face of the crystal, their
curve could  {\it not} represent the true $\rho_{ab}$. Moreover, neither the
current redistribution (discussed in \cite{bush} for homogeneous
medium) nor imperfections of their huge crystals were accounted
for in Ref.\cite{mor}.

\begin{figure}
\begin{center}
\includegraphics[angle=-0,width=0.47\textwidth]{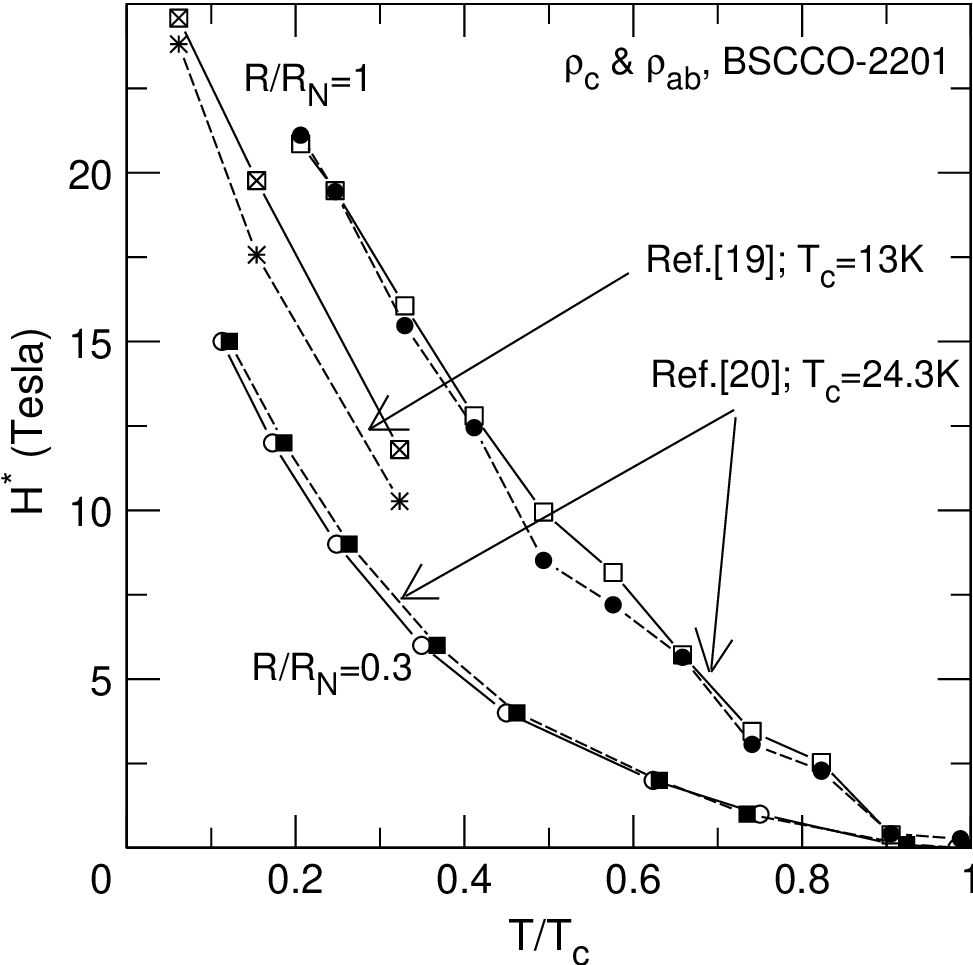}
\vskip -0.5mm \caption{ $H_{c2}(T)$  obtained from independent
resistance measurements in Bi-2201  \cite{ando,zhang}; broken
lines correspond to the data taken from $\rho_c$, solid lines from
 $\rho_{ab}$. }
\end{center}
\end{figure}

 The resistive upper critical field, which is about the same from  in- and out-of-plane
 data for Bi2212,  shows a non-BCS temperature dependence, Fig.1. These results are
 supported by the independent studies of $\rho_c$ and $\rho_{ab}$ in
 a single-layer cuprate Bi2201 with the similar anisotropy.  If we
apply the routine procedure for resistive $H_{c2}(T)$ evaluation
\cite{alezav}, the very similar values of $H_{c2}(T)$ are obtained
from $\rho_{ab}$ $and$ $\rho_c$  measured on the same crystals
\cite{ando} and films \cite{zhang} (see dashed and solid lines in
Fig.2). Remarkably, $H_{c2}(T)$ obtained are compatible with the
Bose-Einstein condensation field of preformed charged
bosons\cite{ale}, and also with some other models \cite{abr,kre}.

Finally we address the origin of the large Nernst voltage measured
above $T_{c0}$ in superconducting cuprates (see  \cite{NERNST} for
more details). It is expressed in terms of the kinetic
coefficients $\sigma _{ij}$ and $\alpha _{ij}$ as \cite{ANSE}
\begin{equation}
e_{y}(T,B)\equiv -{\frac{E_{y}}{{\nabla _{x}T}}}={\frac{{\sigma
_{xx}\alpha _{yx}-\sigma _{yx}\alpha _{xx}}}{{\sigma
_{xx}^{2}+\sigma _{xy}^{2}}}},
\end{equation}
where the current density per spin is given by $j_{i}=\sigma
_{ij}E_{j}+\alpha _{ij}\nabla _{j}T$. Carriers in doped
semiconductors and disordered metals occupy states localised by
disorder and itinerant Bloch-like states. Both types of carriers
contribute to the transport properties, if the chemical potential
$\mu$ (or the Fermi level) is close to the energy, where the
lowest itinerant state appears (i.e. to the mobility edge).
Superconducting cuprates are among such poor conductors and their
superconductivity appears as a result of doping, which inevitably
creates disorder.  Indeed, there is strong experimental evidence
for the coexistence of itinerant and localised carriers in
cuprates in a wide range of doping \cite{tat}.

When the chemical potential is near the mobility edge, and  the
effective mass approximation is applied, there is no Nernst signal
from itinerant carriers alone, because of a so-called Sondheimer
cancellation  \cite{sond}. However, when the localised carriers
contribute to the longitudinal transport, $\sigma _{xx}$ and
$\alpha _{xx}$ in Eq.(1) should be replaced by $\sigma
_{xx}+\sigma _{l}$ and $\alpha _{xx}+\alpha _{l}$, respectively.
Since the Hall mobility of localised carriers is often much
smaller than their drift mobility \cite{mot}, there is no need to
add their contributions to the transverse kinetic coefficients.
One can also neglect field orbital effects because the Hall angle
remains very small for the experimentally accessible fields in
poor conductors, $\Theta_H \ll 1$ \cite {xu,wang1}, so that
\begin{equation}
e_{y}(T,B)={\frac{{{\sigma _{l}\alpha _{yx}-\sigma _{yx}\alpha _{l}}}}{{%
(\sigma _{xx}+\sigma _{l})^{2}}}}.
\end{equation}
 The conductivity of itinerant carriers $\sigma _{xx}$  in the \emph{superconducting }
 cuprates dominates over that  of localised carriers \cite{tat}, $\sigma _{xx}\gg \sigma _{l}$,
 which  simplifies Eq.(2)  as
\begin{equation}
{e_y\over{\rho}}={\frac{k_{B}}{{e}}}r\theta \sigma _{l},
\end{equation}
where $\rho=1/[(2s+1)\sigma_{xx}]$ is the resistivity, $s$ is the
carrier spin, and $r$ is a constant,
\begin{equation}
\frac{r}{2s+1}=\left({e|\alpha_l|\over{k_B\sigma_l}} +
{{\int_{0}^{\infty} dE E(E-\mu)\partial f(E)/\partial E }
\over{k_BT \int_{0}^{\infty} dE E\partial f(E)/\partial E}}\right
)
\end{equation}
Here  $N(E)$ is the density of states (DOS) near the band edge
($E=0$), and $\mu$ is taken with respect to the edge. The ratio
$e|\alpha _{l}|/k_{B}\sigma _{l}$ is a number of the order of one.
For example, $e|\alpha _{l}|/k_{B}\sigma _{l} $$\approx$2.4, if
$\mu$=0 and the conductivity index $\nu$=1 \cite{end}. Calculating
the integrals in Eq.(4)  yields $r$$\approx$14.3 for fermions
($s$=1/2), and $r$$\approx$2.4 for bosons ($s$=0) with the
two-dimensional DOS, $N(E$$)=$$constant$.

 The Nernst signal, Eq.(3), is positive, and its maximum value
 $e_y^{max} \approx (k_B/e)r\Theta$   is about
$5$ to $10$  $\mu$V/K with $\Theta=10^{-2}$ and $\sigma_l \approx
\sigma_{xx}$, as observed \cite{xu,wang1}. Actually, the magnetic
and temperature dependencies of the unusual Nernst effect in
cuprates are described by Eq.(3) quantitatively, if $\sigma_l$
obeys the Mott's law,
\begin{equation}
\sigma_l= \sigma_0 \exp \left[-(T_0/T)^{x}\right],
\end{equation}
where $\sigma_0$ is about a constant. The exponent $x$ depends on
the type of localised wavefunctions and variation of  DOS, $N_l$
below the mobility edge \cite{mot,shk,tok}. In two dimensions one
has $x=1/3$ and $T_0 \approx 8 \alpha^2/(k_BN_l)$, where $N_l$ is
at the Fermi level.

\begin{figure}
\begin{center}
\includegraphics[angle=-0,width=0.47\textwidth]{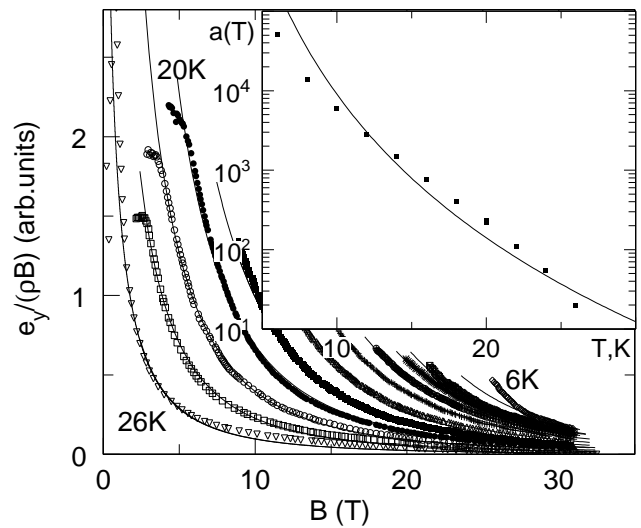}
\vskip -0.5mm \caption{Eq.(6) fits the experimental signal (symbols) in $La_{1.8}Sr_{0.2}CuO_4$ \cite{wang1}
with $b=7.32$(K/Tesla)$^{1/3}$. Inset shows $a(T)$ obtained from the fit (dots)
together with $a$$\propto$$ T^{-6}$ (line). }
\end{center}
\end{figure}

In sufficiently strong magnetic field\cite{ref} the radius of the
'impurity' wave function $\alpha^{-1}$ is about the magnetic
length, $\alpha \approx (eB)^{1/2}$. If the relaxation time of
itinerant carriers is due to the particle-particle collisions, the
Hall angle depends on temperature as  $\Theta_H \propto 1/T^2$, and
the resistivity is linear, since the density of itinerant carriers
is linear in temperature, both for fermionic  and/or bosonic  carriers \cite{alebramot}. Hence, the model explains
the temperature dependence of the normal-state Hall angle and
resistivity in cuprates at sufficiently high  temperatures. Then
using Eq.(3) and Eq.(5) the Nernst signal is given by
\begin{equation}
{e_y\over{B\rho}}= a(T) \exp\left[-b(B/T)^{1/3}\right],
\end{equation}
where $a(T) \propto T^{-2}$ and $b=2[e/(k_BN_l)]^{1/3}$ is a
constant.  The phonon drag effect should be
 taken into account at
low temperatures in any realistic model. Then $a(T)$ in Eq.(6) is
found to be enhanced by this effect as $a(T) \propto T^{-6}$
\cite{NERNST}. The theoretical field dependence of $e_y/(B\rho)$,
Eq.(6),  is in excellent quantitative agreement with the
experiment, as shown in Fig.3 for $b=7.32$ (K/Tesla)$^{1/3}$. The
corresponding temperature dependence of $a(T)$ follows closely
$T^{-6}$, inset to Fig.3. The density of impurity states
$N_l=8e/(b^3k_B)$ is about $4\times 10^{13}$ cm$^{-2}$(eV)$^{-1}$,
which corresponds to the number of impurities  $N_{im}\lesssim
10^{21}$ cm$^{-3}$, as it should be.

If carriers are fermions, then the product $ S\tan\Theta_H$ of the
thermopower $S$ and of the Hall angle  should be larger or of the
same order as $e_y$, because their ratio is proportional to
$\sigma_{xx}/\sigma_{l}\gg 1$ in our model. Although it is the
case in many cuprates, a noticeable suppression of
$S\tan\Theta_H$, as compared with $e_y$, was reported to occur
close to $T_c$ in strongly underdoped LSCO and in a number of
Bi2201 crystals \cite{xu,wang1}. These observation could be
generally understood if we take into account that underdoped
cuprates are strongly correlated systems, so that a substantial
part of carriers is (most probably) preformed bosonic pairs
\cite{alebook}. The second term in Eq.(4) vanishes for (quasi)two
dimensional itinerant bosons, because the denominator diverges
logarithmically if $\mu\approx 0$.  Hence, their contribution to
the thermopower is logarithmically suppressed. It can be almost
cancelled by the opposite sign contribution of the localised
carriers, even if $\sigma_{xx}\gtrsim \sigma_{l}$. When it
happens, the Nernst signal is given by $e_y=\rho \alpha_{xy}$,
where $\alpha_{xy} \propto \tau^2$. Differently from that of
fermions, the relaxation time of bosons is enhanced critically
near the Bose-Einstein condensation temperature, $T_c(B)$,
$\tau\propto[T-T_c(B)]^{-1/2}$, as in atomic Bose-gases
\cite{bec}. Providing $S\tan\Theta_H \ll e_y$, this critical
enhancement of the relaxation time describes well the temperature
dependence of $e_y$ in Bi2201 and in strongly underdoped LSCO
close to $T_c(B)$.

To conclude, we have shown that the understanding of the reliable
experimental data does not require radical revision of the
magnetic phase diagram of cuprates \cite{zavkabale}. Our
studies of $\rho_{ab}(B)$ and $\rho_c(B)$ on the same Bi2212
crystals as well as the normal state model of the Nernst signal in
cuprates support any microscopic theory, which describes the state
above the resistive and magnetic phase transition as perfectly
'normal' with $F(\mathbf{r,r^{\prime }})=0$. The carries could be
normal-state fermions, as in any BCS-like theory of cuprates, or
normal-state  charged bosons, as in the bipolaron theory
\cite{alebook}, or a mixture of both. We believe that the
resistive determinations provide the genuine $H_{c2}(T)$, and the
anomalous Nernst effect in high-$T_{c}$ cuprates is a normal state
phenomenon.

 This work was supported by the Leverhulme Trust (grant F/00261/H).


\begin{thebibliography}{990}
\bibitem{ref} The magnetic field $B\gtrsim 1$ Tesla is strong in
conventional semiconductors with shallow impurity levels
\cite{tok}.
\end{thebibliography}
\end{document}